\pdfoutput=1
\PassOptionsToPackage{table,svgnames}{xcolor}

\documentclass[preprint,hyphens]{vgtc}               
\ifpdf
  \pdfoutput=1\relax                   
  \usepackage{graphicx}                
  \DeclareGraphicsExtensions{.pdf,.png,.jpg,.jpeg} 
\else
  \usepackage{graphicx}                
  \DeclareGraphicsExtensions{.eps}     
\fi%

\graphicspath{{figures/}{pictures/}{images/}{./}} 

\usepackage{microtype}                 
\PassOptionsToPackage{warn}{textcomp}  
\usepackage{textcomp}                  
\usepackage{mathptmx}                  
\usepackage{times}                     
\usepackage{cite}                      
\usepackage{tabu}                      
\usepackage{booktabs}                  

\onlineid{0}

\vgtccategory{Research}

\vgtcinsertpkg

\ieeedoi{10.1109/EduVis60792.2023.00015}

\usepackage{booktabs}                  
\usepackage{tabularx}

\usepackage{mathptmx}                  

\usepackage{multicol}
\usepackage{multirow}
\usepackage{tikz}
\usepackage[noindex,noinline,author=,nomarginclue]{fixme}
\fxusetargetlayout{color}
\fxusetheme{colorsig}
\definecolor{fxtarget}{rgb}{0,0.72,0.92}
\definecolor{fxwarning}{rgb}{0,0.72,0.92}

\usepackage{zref-titleref}
\makeatletter
\newcommand*{\currentname}{\zref@getcurrent{title}}
\makeatother

\definecolor{inroom}{rgb}{0,0,0}
\definecolor{hallway}{rgb}{0.80,0.70,0.70}
\newcommand{\bp}[2]{
\resizebox{20pt}{!}{\begin{tikzpicture}
\draw[inroom, fill] (0,0) rectangle (#1,10);
\draw[hallway, fill] (#1,0) rectangle (#1+#2,10);
\draw[black, line width=40pt] (0,0) rectangle (50,10);
\end{tikzpicture}}}

\newlength{\mylength}
\makeatletter
\newcommand{\mycfs}[1]{%
  \normalsize
  \@defaultunits\mylength=#1pt\relax\@nnil
  \edef\@tempa{{\strip@pt\mylength}}%
  \ifx\protect\@typeset@protect
     \edef\@currsize{\noexpand\mycfs\@tempa}
  \fi
  \mylength=1.2\mylength
  \edef\@tempa{\@tempa{\strip@pt\mylength}}%
  \expandafter\fontsize\@tempa
  \selectfont
}
\makeatother

\hypersetup{breaklinks=true}

\title{Reflections on Designing and Running Visualization Design and Programming Activities in Courses with Many Students}

\author{Søren Knudsen\thanks{e-mail: soekn@itu.dk} %
\and Mathilde Bech Bennetsen
\and Terese Kimmie Høj
\and Camilla Jensen
\and Rebecca Louise Nørskov Jørgensen
\and Christian Søe Loft
}
\affiliation{\scriptsize Digital Design Department\\IT University of Copenhagen\thanks{supplementary materials: \url{https://ituvised.github.io}}}

\abstract{%
In this paper, we reflect on the educational challenges and research opportunities in running data visualization design activities in the context of large courses. 
With the increasing number and sizes of data visualization course, we need to better understand approaches to scaling our teaching efforts. 
We draw on experiences organizing and facilitating activities primarily based on one instance of a master's course given to about 130 students. 
We provide a detailed account of the course with particular focus on the purpose, structure, and outcome of six two-hour design activities.
Based on this, we reflect on three aspects of the course: 
First, how the course scale led us to thoroughly plan, evaluate, and revise communication between students, teaching assistants, and lecturers. 
Second, how we designed learning scaffolds through the design activities, and the reflections we received from students on this matter. 
Finally, we reflect on the diversity of the students that followed the course, the visualization exercises we used, the projects they worked on, and when to key in on simple boring problems and data sets. 
Thus, our paper contributes with discussions about balancing topical diversity, scaling courses to many students, and problem-based learning.%
}


\begin{document}


\firstsection{Introduction}
\maketitle

We reflect on the educational challenges and research opportunities in running visualization design activities in the context of courses with more than hundred students. 
Our work builds primarily on the experiences of activities in a single instance of a master's course at our institution. 
However, this course and our reflections are informed by prior versions of the course, as well as other courses at approximately the same level in the program and at other institutions.
We approach our work through a detailed account of the course and our observations of how various configurations of students and us as educators worked together in the pursuit of learning.

Our work responds to recent discussions in visualization education. Bach et al.~\cite{bach2023challenges} call for a better understanding of motivation during visualization learning experiences, for rigorous examinations of teaching methods and learning activities in visualization education, for  explorations of configurations of teaching environments and how to use them in teaching, and for structuring the materials that we use. 
Specifically, motivated by our duties to facilitate and teach a visualization course attended by about 130 students annually (and increasing), we want to understand good ways to scale successful visualization education methods, thus responding, for example, to how to assess students at scale and to accomodate diversity in doing so~\cite{bach2023challenges}. 

Many of the challenges we face seem perhaps to lack specificity to  visualization in the topical, disciplinary sense.
However, we think that it \textit{is} absolutely visualization specific in the sense of considering the state that visualization education is at---at least, the way that we see it.
Thus, where is visualization education at?
From recent discussions with other visualization educators~\cite{bach_et_al:DagstuhlRep2022}, we find that many educational institutions are creating new visualization courses, many of these rooted in other disciplines such as public health sciences, design schools, and within humanities disciplines.
Additionally, existing courses are growing larger.
While this is anecdotal evidence (in lack of surveys to provide a more rigorous overview), it suggests considering approaches to scale visualization education and dissemination activities more broadly. 
Bach et al.~\cite{bach2023challenges} ask `\textit{how we can provide feedback that is transparent, fair, high-quality, and timely when teaching at scale}'.
Here, we provide an account of our experiences working with 130 student learners from the point of view of the teaching team.
While we think that many our choices and solutions have merit, they are far from perfect.
For example, as course manager, we gave feedback on over four-hundred student reflections---not what we consider a \textit{solution} to the question of scale. 


In this paper, we focus on visualization design activities run as part of a spring 2023 data visualization course at our design program. 
Doing so, we 
%
contribute a rich account of six design activities in terms of purpose, structure, and outcome in addition to accounts of their relation to subsequent programming activities. From this, we present three themes emerging from reflections between members of the teaching team: 
First, scaffolding in our exercise activity design and the relevance of continuous planning.
Second, considerations for balancing homogeneity and heterogeneity in student projects and activities.
Third, based on the first two themes, we reflect on our communication within and between students, teaching assistants, and lecturers. 
Doing so, our reflections form a basis for fellow visualization educators for designing visualization teaching and learning.

\section{Related work}
Here, we introduce related work in visualization education, articulate how visualization design activities build on traditions from design and computer science curricula, 
and finally, summarize research on collaborations between members of a teaching team, such as teaching assistants, lecturers, and course managers.

\subsection{Education in visualization}
Many members of the international visualization research community are actively teaching courses on, or otherwise providing opportunities for, data visualization learning.
One need only look to the undercurrent of recent events organized in relation to premier international venues, such as IEEE VIS, to identify a wealth of education-related topics and discussions.
For example, Vis Pedagogy 2016--17~\cite{WorkshopPedagogy2016,WorkshopPedagogy2017}, VisFutures 2020~\cite{perin2020ieee}, VisActivities 2020--21~\cite{huron2020ieee,huron2021ieee}, and VisGuides 2016, 18, 20, and 22~\cite{VisGuides2016,VisGuides2018,VisGuides2020,VisGuides2022}.
Reading descriptions of these events, the need to teach both computer science and design skills seem to be recurring. 
A similar perspective can be found in visualization textbooks (for example, Munzner~\cite{munzner2014visualization} and Kirk~\cite{kirk2016data}) and recently re-iterated in a call to action for the scholarship of visualization teaching and learning~\cite{bach2023challenges}. 

Perhaps obviously, this is as important from a student perspective as from an educator perspective---after all, all educators have a background as students, and most have primarily been trained within the scope of one discipline.
Thus, visualization education presents a mix of teaching and learning computer science and design skills to students that sometimes arrive with backgrounds in either discipline, or, rarely in our experience, from an application domain discipline.

To better understand and build on these disciplines, we see a need to understand the expectations, knowledge, and culture of activities within these disciplines and how they complement each other.

\subsection{Understanding the interplay of design and computer science education}
Visualization activities as described in visualization research rely extensively on design skills (for example, constructive visualization~\cite{huron2014constructive}).
Yet, to our knowledge, such activities are never said to be rooted in computer science tradition. 
Thus, we first seek to answer what activities (if any) exist within computer science education.

According to a recent overview~\cite{hazzan2020guide}, computer science education has traditionally had limited focus on the activities learners do~\cite{goodyear2021activity} as well as consideration for active learning approaches~\cite{hazzan2020active}.
Yet, active learning approaches, which focuses on what students \textit{do}, are suggested to lead to much improved learning outcomes~\cite{berssanette2021active} compared to traditional approaches.
Recent literature has thus sought to facilitate the move towards active learning in computer science. 
For example, Hazzan et al.~\cite{hazzan2020guide} provide an activity-based guide to teaching computer science.
While practice may be catching up to this movement (this seems to be the case at our institution), the history of computer science teaching may provide a useful backdrop. 

Art and design education seem to stand in stark contrast to the descriptions of computer science.
For example, in an inquiry into four fields in art and design higher education, Shreeve et al.~\cite{shreeve2010kind} describe signature pedagogies.
Particularly relevant for our endavor, they suggest the material and physical dimensions of learning, the visibility of process, and the consideration for teaching and learning spaces.
While art and design programs are likely to have access to studio space, such spaces might be less common elsewhere.
Shreeve et al.~\cite{shreeve2010kind} also discuss the social embedding of teachers and learners, for example, in students participating peripherally~\cite{wenger1998communities} and in students and teachers engaging in communities~\cite{wenger1999communities}.


Understanding these disciplinary traditions, we gain a sense of conflict and mismatch. At least, it seems utterly clear that computer science activities are scalable in a way that does not match the small studio-based learning experienced in art and design education. 
Then, it is worth asking what we are to do with this mismatch?

The approach and thinking we have chosen to follow, is to consider scale in the sense that teaching assistants might be able to take part in responding to this challenge.
While many visualization educators have perhaps considered building communities of practices with students, perhaps we need to start to consider building these communities with teaching assistants?


\subsection{Accounts of teaching team collaboration}
Considering how teaching assistants might fruitfully engage in communities of practice to allow scaling visualization education, we account for research on collaboration in teaching teams.

To our surprise, we did not identify literature that described teaching assistants as members of teaching teams.
Instead, they seem to often be described separately, perhaps explained by their common dual-citizen role as students \textit{and} teachers~\cite{muzaka2009niche}.
For example, Nyquist and Wulff~\cite{nyquist1996working} provide advice for working effectively \textit{with} teaching assistants. 
A further complicating factor, is that there are different cultural expectations for who takes up the role of teaching assistant.
In North America, both our experience as academics as well as literature on the subject~\cite{nyquist1996working} suggest that this role is typically filled by graduate students.
In contrast, this is unusual in a Danish context.
Here, senior students in the study program commonly takes on this role through employment specifically as teaching assistants in fixed-hour contracts.
The combination of role and students' other obligations means that teaching assistants' time resources are typically tightly scoped.
It is our impression that other European countries fit somewhere in this continuum, although postdoctoral scholars might also be considered for this role in some countries. 
We are less aware of the situation in other parts of the world, thus displaying both a lack of literature on this matter, a lack of concrete experience, and by extension, our cultural biases. 

Themes of literature that \textit{do} discuss teaching teams relate to navigating collaboration and cooperation in a team of peer faculty members~\cite{benjamin2000scholarship,koeslag-kreunen2018leadership}.
Here, teaching teams are suggested to allow for sharing and critiquing ideas, and for comfortably engaging in confrontations, where conclusions are confronted in the spirit of deepening understanding about teaching.
Yet, this is rarely happening in practice between faculty who often see teaching teams strictly as a way to distribute work.
Koeslag-Kreunen et al.~\cite{koeslag-kreunen2018leadership} summarizes that members of teaching teams are not used to discuss their work practices with one another and tend to neglect any innovation in their tasks.
However, while the roles and power differences between faculty in a teaching team may be opaque, these differences seem clearer between teaching assistants and lecturers.
Thus, one might take inspiration from these, to consider how more senior members of the teaching team could facilitate shared discussions and development of educational competences. 
Next, we describe the learning context.






 

\section{Learning context}
Here, we introduce the learning context. 
First, we outline the course in relation to its place in a study program and students backgrounds. 
Second, we describe the intended learning ourcomes and assessment of these, framed through the notion of constructive alignment.
Third, we present an overview of the semester, course materials, as well as the teaching team and physical context of the teaching and learning.

\textbf{Study program, student backgrounds, and course history:}
The course was given as a mandatory course in the second semester of a two-year masters' program on `Digital Design and Interactive Technologies'. 
About 110 students were enrolled from this program.
In addition, about 20 students enrolled in the course across our other programs and other local institutions.

The students in our design program came from many different educational backgrounds, including our own bachelor's degrees, as well as backgrounds spanning graphic design, history, and nursing across university and professional school educations.
Almost all students in the study program were new to programming when entering the program.
These students followed an introductory course to programming in Python in the semester preceding our course, while students with limited programming experience, such as through high-school electives and personal interest, enrolled in a course on mobile application development.

In planning and teaching the course, we built on our previous experiences in similar courses.
Most importantly, the course manager had been running the course in the spring 2022 semester.
Here, it was difficult to recruit teaching assistants.
As a result, the five teaching assistants had only limited experience with the course and topic.
In addition, they were challenged by being split across five small 24-people rooms.
Thus, the limited experience with the course and topic were additionally challenged by their inability to support each other in teaching situations.
We thus kept this in mind when preparing for the 2023 iteration of the course.

\textbf{Intended learning outcomes, project work, and assessment:}
In the course, the goal was for students learn how to conceptualize, visualize, and present data with a backdrop of considerations for the potential impact of data and data visualizations. 
The course presented data visualization as an iterative process involving design methods and digital tools to create visualization products, from collecting and organizing data to conceptualizing data visualizations, to the application of these to gain insight.
We designed the course based on the notion of constructive alignment~\cite{biggs2011teaching}. 
Constructive alignment suggests to `\textit{systematically align the teaching/learning activities, as well as the assessment tasks, to the intended learning outcomes}'~\cite[p. 11]{biggs2011teaching}. 
On that note, we next outline learning goals and assessment, while the rest of the paper focus on the activities that students were asked to \textit{do}. 
We framed the goals for the course through intended learning outcomes (ILOs), which we present in Table \ref{tab:intended-learning-outcomes}.


Students worked on a semester-long project in groups of four to five students.
The project asked them to use Altair to create an interactive data visualization from a data set that they collected about their own lives and that they collected during the semester.
At the end of the semester, student groups submitted a report of roughly 8--10,000 words that described their data collection and design process, as well as the visualizations they produced.

In addition to serving as a learning opportunity, the project report served to assess the students learning in relation to the ILOs. 
The assessment was carried out by the lecturer and two external faculty from an examiners corps recruited specifically for this task.
Inspired by Peck's rubrics~\cite{peck2018creating}, we used five rubrics to guide the assessment.
We present these on the supplemental website.

\textbf{Course overview and schedule:}
All in-person course activities were scheduled on Fridays. 
This included teaching team meetings,
teaching assistant work meetings,
lectures,
exercise sessions,
office hours,
and teaching team check-ins.
While teaching assistants attended the first five lectures to ensure cohesion between lectures and exercises, we prioritised teaching assistant time for preparing exercise materials during the later lectures.

We organized eleven exercise sessions about two hours each, scheduled immediately after lectures.
However, to ease students into the semester, we did not schedule an exercise in the first week.
To practice design thinking and reduce the `rush to code', the first five exercises (D1-5) focused on visualization design.
These involved free-hand sketching, constructive visualization, and digital vector-based sketching.
Further, we used these to introduce approaches to developing and refining ideas, such as the 10 plus 10 sketching method~\cite{greenberg2011sketching}.
The next four exercises (A1-4) focused on programming visualizations in Altair (we chose Altair since many students had been introduced to Python in the preceeding semester). 
The second last exercise (D6) focused on visualization evaluation and the last (P1) provided time for project work with support from teaching assistants. 
Table \ref{tab:course-weeks} presents an overview of the semester.

\textbf{Materials and other communication from the teaching team to students:}
For each week of the course, we listed readings via the course page, which served as the main source of course information and digital communication platform.
Students read chapters from \textit{Visualization Analysis \& Design}~\cite{munzner2014visualization}, in addition to selected chapters from other books~\cite{lupi2016dear,kirk2016data,tominski2020interactive,bremer2021data}. 
Additionally, we provided suggestions for papers to read with relevance to the lecture topics.

We prepared a slide deck for all design exercise sessions (D1--6) and for all but the last Altair exercise session (A1--3). 
The slide decks were made available at the start of the exercise via the course page on our institution's learning management system. 
In addition to materials specific to that day's exercises, all slide decks outlined the plan for the exercise session.
Further, it reminded students to capture process documentation for their reflection hand-ins and project process, for example, by taking pictures.

During in-person course days, we drafted an e-mail to share with students on the course page and relayed to their e-mail accounts at the end of the day.
This summarized the days' course activities, gave clarifications based on questions we received during lectures, exercises, and office hours, and reiterated the plan for the next weeks.

\begin{table}[tb!]
    \caption{Course intended learning outcomes (ILOs)~\cite{biggs2011teaching}. }
    \label{tab:intended-learning-outcomes}
    \centering
	\setlength\tabcolsep{0.10cm}
	\def\arraystretch{0.99}
    \mycfs{7.3}
 \begin{tabularx}{\linewidth}{@{} l X @{}}
    \hline
\\[-6pt]
ILO 1 & Sketch novel data visualization designs and build interactive visualization prototypes.\\
ILO 2 & Explain fundamental theories and design principles in data visualization, apply them in a design process, and reflect on these.\\
ILO 3 & Interpret, deconstruct, and critique data visualizations.\\
ILO 4 & Reflect on the ethical and societal implications of data visualization.\\
\\[-6pt]
    \hline
    \end{tabularx}
    \vspace{-0.4cm}
\end{table}

\textbf{Student submissions and other communication from students to the teaching team:}
During the course, students could submit six design exercise reflections and a project design brief.
To submit their projects as an exam submission, students needed to have three design exercise hand-ins approved.
These were brief 400-word essays that asked students to describe what they did during the exercise, to reflect on the process and on the result. 
We intentionally under-specified these reflection pieces, as we wanted students to decide for themselves what they found important to reflect upon.
Based on the limited student presence in the prior course iteration, we implemented these hand-ins to emphasize the exercise importance to students.
This echoes the suggestion by Biggs and Tang to require `\textit{students to engage in the learning activities required in the outcomes}'~\cite{biggs2011teaching}.
While formally these were possible to fail, we passed any submission that showed concerted effort to attend the exercise and reflect on it.
More importantly though, we provided individual written feedback to almost all reflection pieces.
The design brief worked as a planning document for students to describe their expected data collection process and their design goals.
While this was not a mandatory submission, all student groups submitted a design brief.
Similar to the reflection pieces, we provided individual written feedback on the design briefs.
These most often related to the project focus, such as the type of data the group planned to collect.
For example, we suggested visualization techniques to consider, visualization exemplars to consult, specific data collection and processing approaches (such as Knudsen et al.~\cite{knudsen2016using}), and potential pitfalls (such as Sedlmair et al.~\cite{sedlmair2012design}).
Inspired from the `not-yet' growth mindset and specifications grading~\cite{nilson2023specifications}, we suggested about a third of the groups to revise and resubmit their design brief.

\begin{table*}[t!]
    \caption{Overview of course weeks and exercise sessions. We include exercise duration (semester project time is shown in \textit{cursive}), student presence, and number of hand-ins.
    The student presence estimations are indicated in a stacked bar (range: 0--50). Black: students that remained in the room after our exercise introduction. Gray: students that worked in the hallway near the exercise room after our exercise introduction.
    }
    \label{tab:course-weeks}
    \centering
	\setlength\tabcolsep{0.10cm}
	\def\arraystretch{0.99}
    \mycfs{7.3}
 \begin{tabularx}{\linewidth}{@{} l l X l p{4cm} p{1.2cm} l l l r @{}}
    \hline
\\[-6pt]
    \textbf{Date} & 
        \textbf{Week} & 
            \textbf{Lecture topic} & 
                \textbf{Exercise} & 
                    \textbf{Exercise topic and inspiration} & 
                        \multirow{2}{1.2cm}{\textbf{Duration (minutes)}} & 
                            \multicolumn{3}{l}{\textbf{Students present (est.)}} &
                                            \multirow{2}{1.1cm}{\textbf{Handins / CIQ's}}
                                                                        \\
    &   &   &   &   &   &    \textbf{Room 1} &
                                \textbf{Room 2} &
                                    \textbf{Room 3} & \\
\\[-6pt] \hline \\[-6pt]
    Feb 10 & \textbf{ 1} & Course overview; What is visualization and why should we bother? \\
\\[-6pt] \hline \\[-6pt]
    Feb 17 & \textbf{ 2} & Deconstructing visualization: Data abstraction, data marks, and channels &  \textbf{D1} & Critique and redesign                                      & 45+15+45   & 
                \bp{30}{0} & \bp{30}{0} & \bp{30}{0} &  119 / 1    \\
    Feb 24 & \textbf{ 3} & Overview of visualization design processes and working with data         &  \textbf{D2} & Data collection and sketching, design brief                              & 45+15, \textit{45} & 
                \bp{30}{0} & \bp{30}{0} & \bp{25}{0} &  115 /12   \\
    Mar  3 & \textbf{ 4} & Understanding user goals and tasks                                       &  \textbf{D3} & Constructive visualization~\cite{huron2014constructive}    & 55+15+60   & 
                \bp{15}{15} & \bp{5}{20} & \bp{5}{20} &  110 / 8    \\
    Mar 10 & \textbf{ 5} & Creating the design                                                      &  \textbf{D4} & Refining visualization ideas~\cite{greenberg2011sketching}                & 45+15+45   & 
                \bp{25}{0} & \bp{15}{7} & \bp{15}{8} &  52 / 4    \\
    Mar 17 & \textbf{ 6} & Creating the design: Going digital                                       &  \textbf{D5} & Sketching in the digital medium                            & 45+15+45   & 
                \bp{25}{0} & \bp{15}{7} & \bp{15}{8} &  18 / 3    \\
\\[-6pt] \hline \\[-6pt]
    Mar 24 & \textbf{ 7} & Introduction to Altair and using it in design                            &  \textbf{A1} & Intro to Altair~\cite{dork2023data}                                            & 45+15+45   & 
                \bp{30}{0} & \bp{25}{0} & \bp{30}{0} &  - / 2 \\
    Mar 31 & \textbf{ 8} & Visualization theory in Altair                                           &  \textbf{A2} & Domains and scales in Altair                               & 45+15+45   & 
                \bp{30}{0} & \bp{10}{20} & \bp{20}{5} &  - / 0 \\
    Apr 14 & \textbf{ 9} & Multiple views and transformations in Altair                             &  \textbf{A3} & Faceting and multiple views in Altair                      & 45+15+45   & 
                \bp{6}{14} & \bp{15}{0} & \bp{20}{5} &  - / 1 \\
    Apr 21 & \textbf{10} & Interaction and interaction in Altair                                    &  \textbf{A4} & Specifying interaction in Altair                           & 45+15+45   & 
                \bp{6}{14} & \bp{15}{8} & \bp{25}{0} &  - / 0 \\
\\[-6pt] \hline \\[-6pt]
    Apr 28 & \textbf{11} & Evaluating and communicating the design                                  &  \textbf{D6} & Heuristic evaluation \cite{zuk2006heuristics}                                      & 45+15, \textit{45}   & 
                \bp{30}{20} & \bp{00}{20} & \bp{0}{20} &  11 / 0 \\
    May 12 & \textbf{12} & Recap, evaluation, perspectives, and questions                           &  \textbf{P1} & \textit{Project work}                                      & \textit{105}   & 
                \bp{20}{30} & \bp{0}{30} & \bp{0}{30} &  - / - \\
\\[-6pt] 
\hline
\hline
\\[-6pt] 
    May 17 & \textbf{13} & \multicolumn{8}{l}{\textit{Semester project deadline. Submissions: 29 group projects across 129 students (14 groups of five, 12 of four, 1 of three, two and one student)}}  \\
\\[-6pt] \hline
    \end{tabularx}
    \vspace{-0.25cm}
\end{table*}

Additionally, we invited students to submit weekly anonymous formative evaluations.
The style and wording of these were inspired by the Critical Incident Questionnaire (CIQ)~\cite{brookfield1998critically}. 
In contrast to descriptions of CIQ, we used a digital submission system on the course page due to the potential number of submissions.


\textbf{Teaching team:}
\label{sec:learning-context}
The teaching team comprised seven people: A lecturer/course manager and six teaching assistants, five of which are authors (the sixth teaching assistant is recognized in acknowledgments).
As we noted above, teaching assistants in Denmark are most commonly more senior students in the study program.
This was also the case for the six teaching assistants in this course, who were employed through 70-hour fixed contracts.
Luckily, we were able to compose a strong teaching team that already knew each other and the lecturer well.
All teaching assistants had attended a course on a specialized topic given by the course manager in the semester prior to our visualization course.
And they had all followed the prior iteration of the course that we focus on in this paper.
Thus, we were in a good state to collaborate with each other.

Yet, the six teaching assistants had limited experience as students in similar design and programming exercises as the previous course iteration had been challenging to run with a large student cohort, inexperienced teaching assistants, and inappropriate physical learning context.
Additionally, since the teaching assistants' time was reserved for collaborative preparation and meeting students in the exercise hours, they had limited opportunities, for example, to read course literature.
Thus, in some respects, we were paving the road as we went---we aimed to stay one step ahead of the students following the course.
For example, when we commenced on the course and semester, we had a goal that all exercise materials should be ready for our teaching meeting one week in advance of using them in the exercises.
We met this goal until about four weeks into the course when our hopes hit reality.
From then on, exercise materials were ready at about an hour before they were needed.


\textbf{The physical learning context:}
With our prior experience of room challenges, we had made sure to receive room assignments that would support the exercise activities we had planned.
We thus asked for, and were assigned three rooms in the same building, each occupying up to fifty people.
Two rooms (Room 1 and 2) were on the second floor, while the third room (Room 3) was directly above Room 1 on the floor above.
In addition, we knew that students rarely made their way to the course manager's office.
Thus, we decided to book a nearby meeting room 
the entire in-person day, which facilitated planning and work meetings in the teaching team.
More importantly however, the room also facilitated `office hours' that were conveniently located directly in front of the exercise rooms.

Next, we describe each exercise in chronological order by way of its purpose, our preparations for it, and the observed outcomes.
We provide additional details on our supplemental website.


\subsection{Design exercise 1: Critique and redesign}
The first design exercise asked students to deconstruct, critique, and redesign data visualizations.

\textbf{Purpose:}
We wanted to provide students with an overview of the topic of data visualization and what concrete elements might be in play.
In addition, since this was the first time students worked with visualizations, we wanted to students to experience how the same data can be represented in many different ways. 

\textbf{Preparation, materials, and structure:}
Before the exercise, we identified five recent data visualizations published on online news sites. 
To motivate students to learn visualization, the exercise asked students to consider both representational aspects and the context in which visualizations appear. 
With this in mind, we based the exercise on visualizations related to climate change (4) or the recent earthquake in Turkey and Syria (1). 
We provided the five visualizations to students in print (A4 sheets) and in the exercise slide deck, the latter with a brief description of the visualizations, and a link to their original context.
The slide deck enabled the students to understand the context in which the visualizations appeared and any form of interactivity.

During the exercise session, the students worked in groups of three to five.
In their group, they were instructed to pick two of the five data visualizations for their work.
First, they were asked to spend up to 15 minutes to analyze the data visualizations. 
We instructed the students to gain an overview of which data the visualizations were based upon. 
Afterwards they were instructed to spend up to 30 minutes to redesign the data visualizations using pen and paper. 
We presented the redesign task as an open task.
Thus, students could freely choose from simple surface-level redesigns (such as  color changes), to fully revise a smaller part of the visualization, to a complete redesign of the entire visualization, including using another visualization type.
At the end of the exercise, we invited students to put their sketches on a whiteboard, to share and discuss their work in plenum.

\textbf{Outcome:}
The exercise worked as an introduction to how data can be represented in different ways, and that it is important to strike a balance between simplicity and complexity---that complex data visualizations might look impressive but simpler design choices might be more appropriate for the given design problem. 
For example, some students stated in the discussion how seeing the same data visualized differently gave them a sense of the visualization designer's responsibility in visualizing data. 
This led to a criticism of the complexity of the earthquake visualization.
Students were concerned that the visual complexity challenged the purpose of communicating the data and visualization broadly, including to people with  limited visualization literacy. 

However, while the five visualizations provided a strong anchoring in current societal issues, we think more thought to the type of visualization and its quality might be warranted.
On the one hand, imperfect examples allow students to sense a need to think critically about visualization design.
On the other, there is also value in showing visualization exemplars early on.
Additionally, at this point in our course, we wonder if it is sensible to provide a broad perspective on what visualization might be, or to rather keep to what they might realistically produce given the tools we introduce during the course.


\subsection{Design exercise 2: Data collection and sketching}
The second design exercise asked students to identify and organize data that was already collected about themselves and to then sketch visualization of the data.
We divided the exercise session in two sections of 45 minutes. The first focused on data collection and pen-and-paper sketching, while the latter focused on student projects. 

\textbf{Purpose:}
We wanted to give students a sense of different opportunities for collecting data about themselves as well as how to externalize their data visualization ideas by sketching in hand.
Secondly, the exercise came at a time when groups of students were starting to discuss the topic of their personal data visualization projects. 
Thus, the exercise served to kickstart this process by getting students to think about their projects, and to discuss it in their groups. 

We used the knowledge students had gained from the first design exercise as scaffolds. 
Most importantly,
it relied on pen-and-paper data sketching. 
Additionally, it built on theoretical aspects of data visualization introduced during lectures and readings.
In contrast to the last exercise, this exercise focused on data about themselves.
The students were asked to identify pre-existing and -collected data about themselves, to organize it for themselves, and then to sketch data visualizations of this data. 

\textbf{Preparation, materials, and structure:}
Before the exercise session, we prepared a slide deck with inspiration for finding data that might already exist about themselves.
For example, we suggested to look at their phones for photo meta-data, message histories,
step counts, screentime data, and media use.
The slide deck also prompted students to work on the project brief. 

In the first half of the exercise session, we asked students to work with individual datasets independently from their groups.
However, we encouraged them to sit near each other, to enable them to inspire each other, to reflect on their ideas, and to discuss their data when needed. 
When students had identified an interesting dataset, they were prompted to reflect on their data, how it might be visualized, and to start externalizing their visualization ideas in pen-and-paper sketches.
We encouraged them to be creative and not be limited by what they had seen or knew already (such as bar charts).

In the second half of the exercise session, we asked students to discuss their project ideas, possibly inspired by lecture examples or the first part of the exercise.
Since they were asked to collect and work with personal data, we also asked them to consider, discuss, and clarify how they felt about sharing their data within the group and more broadly (other students, teaching team, and examination team).
Concretely, they were asked to balance the personal data aspect with working with data that they found interesting and doable. 
With `interesting', we suggested something that might allow them to gain new insights about themselves.
With `doable', we suggested how they would be able to capture the data and the time needed for data collection (both work hours and calendar time).




\textbf{Outcome:}
The exercise worked to help students think about their project work.
We think students benefitted from initially working independently along the other group members.
This allowed them to sketch their own ideas while also being able to discuss their thoughts and ideas with other group members.
Additionally, students' independent examination of data sources about themselves 
provided a solid framing for the groups to discuss ideas for their design brief. 
However, we also talked to and observed students who experienced challenges in working creatively with data visualization ideas at this point in the course, likely due to limited exposure to existing data visualization examples and visualization theory.


\subsection{Design exercise 3: Constructive visualization}
The third design exercise asked students to use Lego bricks to collaboratively create physical data representations and to evaluate these with their peers. 
We based the exercise on the Constructive VizKit~\cite{huron2016using,willett2016constructive} and the activities described as part of it.

\textbf{Purpose:}
We wanted to give students a sense of alternative visualization materialities and their potential role in visualization design. 
Most importantly, with the co-constructive nature of building visualizations with Lego, we intended for students to experience building off each other's ideas and to do quick experiments in the process of creating, destroying, and building something anew. 
Hopefully, to then gain a sense of working with the creative design process.

\textbf{Preparation, materials, and structure:}
Before the exercise session, we assembled materials with inspiration from visualization education literature~\cite{willett2016constructive,huron2016using}. 
%
%
In addition to suggestions from the literature, we considered how the materials would be distributed to the many students, how it would be collected from them again, and how it could be stored and ready for new use.
While we experimented with creating custom boxes of various materials (acrylic sheets, fiberboard, foam boards), at the end, we decided to order transparent boxes in various sizes, with the aim to find a box that would: 1) snugly fit a set of bricks, 2) make it easy for students to use and store bricks, and 3) make it easy for us to distribute, receive, and store without losing parts.
We settled on a transparent acrylic box that snugly fit 32 bricks in 8 different colors (slightly less than suggested elsewhere~\cite{willett2016constructive}) and compiled 30 sets of this configuration.
Finally, we used exercise sheets provided by Huron et al.~\cite{huron2016using} and prepared a slide deck with inspiration from these materials.

Initially, we introduced students to the constructive exercise (15 min). Then we handed out data sheets and let students start to work with the data sheets, discussing, analyzing, and building (40 min). We prompted the students to try as many different visualizations as possible. Further, we suggested the use of additional auxiliary materials. 
For example, to consider using sticky notes for creating legends and axes or other annotations.
We also specifically requested them to not open their laptop for this task.
After a break (20 min), we distributed self-assessment sheets with a brief persona description corresponding to their data set plus five two-part questions. 
First, we asked students to read the self-assessment sheet and consider the questions in relation to their visualization.
Then, they were asked to discuss this question in their group and to consider re-designs.
After doing the self-assessment (20 min), we distributed peer assessment sheets
and asked students to read them.
Based on the sheets, we asked each group to split in two: 
One half stayed with their construction to facilitate a brief evaluation based on the peer assessment sheet.
The other half visited other groups as evaluation participants (35 min). 
Doing so, we asked students to consider what personas the other groups had and what they could deduce from their visualizations.
At the end of the session, we reconvened in plenum to briefly summarize the learning points from the exercise (5 min).


\textbf{Outcome:}
The exercise worked to make the students aware of the benefits of exploring a wide range of visualization design ideas.
In contrast to earlier pen-and-paper based sketching, the exercise made it much easier for the students to construct and tear apart ideas, to deconstruct ideas.
Students experienced how they could use the bricks as malleable materials that provided a stronger connection between data collection, data wrangling, and data representation. 
Particularly the collaborative nature of the building blocks allowed students to discuss different representational ideas within their group.



\subsection{Design exercise 4: Refining visualization ideas}
The fourth design exercise asked students to use the 10 plus 10 sketching method~\cite{greenberg2011sketching} for a predefined data set as well as their own. Here student groups were to create a series of ten sketches, each focusing on a specific aspect of the dataset.
Then, they would select one sketch to explore in 10 further sketches.

\textbf{Purpose:}
We wanted to provide concrete approaches for students work with ideas.
We facilitated this through experimenting with quick sketching within a time limit.
Second, the repetition of the sketching process itself promoted familiarity and fluidity, facilitating a deeper understanding of divergent and convergent thinking.

\textbf{Preparation, materials, and structure:}
We divided the design exercise in two. In the first half, the students were instructed to apply the 10 plus 10 sketching method to an existing dataset,
which offered a wide range of sketching opportunities.
%
In the other half, students were expected to use 10 plus 10 with their own dataset. Thus, after completing the initial set of sketches, each group member assessed their own work and selected one sketch that they deemed the best or most interesting among their creations. Building upon the chosen sketch, each group member then went on to the second phase of the exercise, where they created ten new sketches based on their selected piece. We encouraged student to discuss and be inspired by one's own or group members sketches to promote an iterative mindset. 

\textbf{Outcome:}
The exercise enabled students to gain more familiarity with their own datasets and exercised their ability to quickly create sketches showing the most important aspects of their visualization ideas. Further, the iterative approach of the 10 plus 10 method made the students revisit and expand upon their initial ideas.
Thus, the exercise enabled them to explore alternative approaches, variations, and potential improvements, which ultimately helped to identify ideas that could be used in their projects.

However, the teaching team had different operationalisations of the 10 plus 10 approach.
While the idea is to mimic divergent and convergent thinking in the two parts (first seek ten different ideas, and then explore, improve, or perfect one of these in ten new variations), some students went away from the exercise with ten explorations of the data  we had provided and ten of their own data. 


\subsection{Design exercise 5: Sketching in the digital medium}
The fifth design exercise asked students to consider the transition to digital tools, specifically RAWGraphs and Figma, for data exploration and visualization.

\textbf{Purpose:}
The primary objective of this activity was to leverage the capabilities of RAWGraphs and Figma to enhance students' understanding of data visualization and its digital applications. We aimed for students to create visually compelling and informative data visualizations using their own datasets or pre-existing datasets.

\textbf{Preparation, materials, and structure:}
The activity began with the creation of a csv file containing relevant data, either from the students' project datasets or a pre-existing dataset. The students then imported the dataset into RAWGraphs, an online tool designed for data visualization. They explored a variety of visualization types offered by RAWGraphs, experimenting with different techniques to identify the most appropriate representations for their datasets.

Once the visualizations were finalized in RAWGraphs, the students exported them as svg files and transitioned to Figma, a versatile digital design tool. Drawing inspiration from previous sketches and concepts, they reimagined and redesigned the visualizations in Figma. With Figma's extensive design capabilities, the students refined visual elements, adjusted layouts, and incorporated additional design elements to enhance both aesthetic and functional aspects.

Following the design phase, the students engaged in group discussions to reflect the advantages and disadvantages of using Figma and RAWGraphs. They compared the efficiency of digital tools with traditional hand-sketching methods from previous week’s activities. Furthermore, they discussed the distinctions between Figma and RAWGraphs, with Figma offering greater flexibility and customization options while RAWGraphs provided a specialized environment for quick exploration of visualization types.

Students considered limitations and possibilities of the tools in group discussions. RAWGraphs stood out for its ease of use and predefined visualization types but posed limitations with regard to customization and handling complex data. On the other hand, Figma's extensive capabilities came with a steep learning curve.

\textbf{Outcome:}
Through this design activity, the students gained practical experience with digital tools for data visualization, expanding their understanding of the benefits and drawbacks of RAWGraphs and Figma. By utilizing these tools, they enhanced their ability to create visually appealing and informative visualizations while considering the limitations and possibilities of each tool.


\subsection{Programming exercises}
After the first five design exercises in the course, we pivoted with a four-week deep dive into programming visualizations with Altair.
Over the four exercise classes (roughly two hours each), we introduced core aspects of Altair to enable the students to use this technology in their project work. 
A table on the supplemental website presents an overview of the four exercises and their focus.

For most of the students Altair was a new and unfamiliar programming tool.
Thus, we planned the exercises such that each were based on the coding artifacts and concepts of exercises before it.
We did so to ensure a smooth and effective introduction to working with data using a programming tool that would not overwhelm the students. 
Additionally, the programming activities followed lectures that introduced related visualization theory. 
For example, for the third exercise (faceting and multiple views in Altair), the topic of the lecture was on multiple views and dashboard layouts.

The purpose of introducing a visualization library was (of course) to enable the students to create interactive, computer-supported data visualizations, which was both part of the course intended learning outcomes and necessary for students to learn to be able to hand in a successful project.
We decided on using Altair for mainly two reasons. 
First, most students followed an introduction to programming course based on Python in the previous semester.
Second, since Altair is based on the notion of a visualization grammar, it nicely supports learning visualization theoretic concepts, such as visual variables and their efficiency.

During live coding sessions, students were introduced to Google Colab, shown how to import csv files and create Pandas dataframes in this environment, and introduced to basic Altair visualization examples. 
Then, students were instructed to attempt these tasks on their own.
To facilitate this, we provided a skeleton Python notebook with alternating Markdown task cells and cells with prompts to `\# Write your code here'. 
We took inspiration from similar course taught elsewhere (for example, at FH Potsdam by Dörk ~\cite{dork2023data}).

We used the cereal data set both during our live coding and with the initial self-directed tasks.
The cereal dataset is simple and contain information on different cereal products. 
We chose to use this data set because it is sufficiently large to enable visualization demonstrations that represent more data than would be possible to manually gain an overview of, while the semantics of the data are easy to grasp, and it contains a useful variety of data types.
We expected these qualities would help students get started with Altair.

\paragraph{Progression of Altair exercises}
Through the four Altair exercises, students were introduced to core Altair concepts through hands-on work, such as domains and scales, faceting and multiple views, and specifying interaction. 
Our goal with the exercises were to push students to get their hands dirty with Altair programming, rather than necessarily getting to the end of the skeleton notebooks.
We estimate that about half of the students had worked their way through the tasks outlined in the skeleton notebooks during the exercise session or before the exercise in the following week.

\textbf{The first Altair exercise} introduced students to the fundamentals of Altair, enabling them to understand its core concepts and syntax. 
This exercise laid the foundation for their journey into the realm of data visualization using Altair. 
At this point in the course, not all groups had gathered sufficient data for their project work. 
Thus, to ensure all students were able to do the exercise, we decided to base it on the cereal data set, which is available from the Vega project. 
We spent half of the exercise session on introducing the tools and live coding (45 min) and devoted the second half to self-paced exercises based on a skeleton notebook (presented in tabular form on the supplemental website).

\textbf{The second Altair exercise} introduced students to domain and scales, and thus served to bridge theory of visualization variables with practice.
Since many groups had not yet gathered sufficient data for their project work, we continued to use the cereal data set.
Additionally, while we had planned to introduce new aspects of Altair, our experience was that we needed to answer many questions similar to the first exercise, which obviously hindered students' ability to move forward at the same pace.
We thus went faster through live coding in this exercise (25 min) and reserved the remaining time (20+45 mins) for students to work on their own.
This also allowed the few groups that were ready to work with their own data to do so.

\textbf{Mid-way teaching team reflections:} We had planned for all students to be able to solve basic tasks in Pandas and Altair after the first two Altair exercises.
However, we found that students were at different levels.
From our interactions with students, we think a variety of factors played into this: Some groups had focused more on collecting data and had run into issues that needed fixing in relation to this. 
For this reason, they had paid less attention to the exercise plan.
Other groups ran into practical or technical issues.
For example, several groups had issues working collaboratively with their data and importing it into Colab.

Since students said that the live coding introductions were useful, we decided to keep these for the remaining Altair exercise sessions. 
However, due to the varied state of students' projects, we kept them brief. 
While students expressed the usefulness of them, many experienced hitting a wall when they needed to work on their own and needed our help to move forward.
Thus, with brief live coding, we had more time for hands-on assistance when students got stuck.

\textbf{The third Altair exercise} introduced students to faceting and multiple views.
This topic closely followed the preceding lecture as well as the structure of the second Altair exercise (first live coding, then independent work).
Since many groups had collected sufficient data for their projects to be able to work on the project at the time of this exercise, most groups used their own data for the exercise and only a handful of groups used the cereal data set.

\textbf{The fourth and final Altair} exercise introduced students to specifying interaction in Altair.
Again, this topic closely followed the preceding lecture as well as the structure of the other Altair exercises. 
To our knowledge, only one or two groups were not ready to work with their project data during this exercise. 

\textbf{At the end of the four Altair exercises}, we had provided an introduction to Altair that would allow students to create visualizations appropriate for their projects. 


\subsection{Design exercise 6: Heuristic evaluation}
The last design exercise asked student groups to perform a heuristic evaluation based on their own project and to team up with another group to do the same with them. 
We divided the exercise session in two sections of 45 minutes, the first focused on heuristics, while the latter provided time for project work.

\textbf{Purpose:}
We wanted the last exercise to provide students with a capstone of their venture into the topic of data visualization and their project work.
In addition, since this was the last exercise, we wanted to help students connect elements of the course to their previous, parallel, and future studies in their digital design study program. 

\textbf{Preparation, materials, and structure:}
Before the exercise, we briefly examined literature on heuristic evaluation in visualization.
Based on this, we decided to use what we think of as an early description of heuristic evaluation framed in terms of evaluating visualizations~\cite{zuk2006heuristics}.
To help students understand the relevance of the exercise and to connect it to other parts of the course, we suggested their insights could be used to explain and discuss the design choices in their project submissions.
The slide deck provided an overview table of the heuristics discussed by Zuk et al.~\cite{zuk2006heuristics} in addition to the common plan for the exercise session.

In the first half of the exercise session, students focused on the heuristic evaluation. 
First, we introduced the exercise plan and facilitated within-group discussions to decide on two heuristics to use for the evaluation, in addition to three heuristics that we suggested (10 min).
We asked students to take inspiration from Zuk et al.~\cite{zuk2006heuristics} as well as discussions throughout the course, such as Munzner's \textit{Rules of thumb}~\cite{munzner2014visualization} or the `Tuftean perspective' (e.g., \cite{tufte2001visual}).
Then, we asked them to evaluate their visualization products using the heuristics they had chosen within their groups (15 min).
Afterwards, they were instructed to team up with another group and use the heuristics to evaluate their own work and that of the other group (20 min).
Thus, students could freely choose from heuristics that focused on, for example, visual variables and other perceptual aspects of visualization, visualization tasks or other cognitive aspects of visualization, or on aspects of visual and graphic design. 
In the second half of the exercise session, we suggested they continued their project work. 
Thus, we did not do a plenary summary at the end of this exercise.

\textbf{Outcome:}
While we think the exercise worked as we had intended for the students that followed it, we estimate that perhaps only 10\% of the students in the course did so.
This was likely a combination of end-of-semester stress and poor timing of the exercise in relation to the students' progress on their project work, which meant that many groups were simply not yet at a stage where they could meaningfully perform any kind of evaluation.

\section{Approach}

We take a reflective autoethnographic approach to report on our experiences during teaching a recent large visualization course. 
%
We met in the week before the design activity to plan the activity. 
During this time, we also reflected on our experiences with prior activities to inform the next ones. 
Additionally, we met briefly after each activity to share experiences and check in with each other. 

After running all six activities, we met twice to reflect in a more structured manner. 
Our first meeting was a brief one-hour meeting.
We used this meeting to discuss and take note of things we found relevant or necessary to record in writing.

Our second meeting occurred about a month after the last exercise session and comprised a more in-depth reflection and analysis over roughly seven hours.
During this time, we split up to reflect individually and in writing, on activities and our individual teaching context (that is, room, students, self).
%
%
Next, we analyzed our written reflection memos.
In the next section, we present the synthesis of our reflection notes according to three themes.

\section{Themes emerging from our reflections}
\label{sec:themes}
In the following, we present the three themes that emerged from our reflections.
First, we discuss how we designed the exercise activities for learning scaffolding and the relevance of continuous planning and reflection in relation to this.
Second, we discuss diversity in exercises and projects, which we frame as seeking a balance between homogeneity and heterogeneity.
Third, based on the first two themes, we discuss the communication that happened within and between students, teaching assistants, and lecturers.

\subsection{Continous planning for scaffolded learning}
Scaffolding is an important concept in learning, which means to create opportunities for learning that leverages existing skills and knowledge.
In our course, scaffolding was continuously at odds with allowing diversity in student work (such as asking students to work with their own data sets) and the need for proper and early planning. 
We initially set a goal to be fully prepared for exercises a week in advance of the exercise, this ambition fell flat (as described in Section \ref{sec:learning-context}).
An important part of why this failed, is that we needed to continuously adapt and plan according to how students reacted to the learning contexts and opportunities.

\textbf{Scaffolding in transitions of skill, method, and material:}
One aspect of scaffolding in our course experiences relate to the transition from pen-and-paper-sketching, to digital vector-based sketching, to programming visualizations.

We structured our visualization course according to visualization design process.
First, we introduced pen-and-paper visualization sketching and constructive visualization concepts.
Second, we moved towards digital sketching and experimented with integrating RAWgraphs into the workflow.
Finally, we introduced tools for creating visualizations from program code.
Students seemed to experience much stronger challenges at each of these transitions.

Reflecting on these observations, we realize that aspects of this seems almost obviously difficult.
Consulting visualization literature, we can think of only three sources that provide concrete guidance on how to approach these transitions in design work \cite{lupi2016dear,bremer2021data,huron2022making}.

From the perspective of motivation, we think there is a lot of value in these shifts of course focus through the semester.
Particularly, the transition from sketching to programming seemed to cause renewed energy among students.
While they mid-way through the course had got used to 
data sketching, they suddenly experienced eureka moments, when getting their program code to produce visualizations.

Since our students predominantly commenced on visualization learning from a design background, the transition difficulties mainly seemed to arise from a lack of technical skills.
However, we wonder if other student cohorts might find design transitions more difficult to grasp, such as transitioning from divergent to convergent thinking, for example, according to the double diamond design process model.

Yet, we consider alternative approaches to structuring the course.
In keeping with the general structure, we consider that a flip-flop model between drawing and programming may provide benefits in terms of scaffolding.
Other potential course structures could abandon the visualization design process structure in favor of structuring by the visualization pipeline, the nested model of visualization design and validation, or something else entirely.
What stands is the need to consider scaffolding in designing any such structure, such that students can follow the learning points, and to consider extended support for scaffolding through transitions, such as going from hand- to vector-based drawing, or from drawing to programming.

We identify the following \textbf{course design goals for scaffolding:}
\textbf{Goal-1:}
Aim for continuous, iterative, course planning and design to ensure proper scaffolding for both students and teaching team. 
Although we note that it is important to ensure alignment of the teaching team in terms of communication, priorities, and available resources. 
\textbf{Goal-2:}
Pay particular attention to scaffolding when designing course transitions, such as going from pen-and-paper-sketching, to digital vector-based sketching, to programming visualizations.
\textbf{Goal-3:}
Include considerations for student diversity when designing for scaffolding, such as students from different educational backgrounds in- and outside computer science or design. 

\subsection{Balancing homo- and heterogeneity in problem-based learning and activity design}
Learners' agency of their own learning is an important factor in motivation.
Thus, we strive to give students agency to decide what they are learning and how.
Here, we frame such choice with the notion of supporting a heterogenous set of problems.

We considered these aspects thoughout the course.
For example, we provided a set of five visualizations to choose from when we asked students to deconstruct existing visualizations in D1.
And in D2, we asked students to work from data they identified on their own.
Yet, our considerations also frequently led us to choose for the students.
For example, D3 relied on data published as part of the public exercise materials.
And when we transitioned to Altair, we used the same cereal data set week after week.

When we decided which data students should work from, or the visualizations they should look at, our goal was for students to spend their time on what we considered most relevant in terms of learning outcome.
Yet, in some situations, it seemed fastest and most in line with the goals to let them work on self-defined data (as in D2). 

\textbf{Data constancy emphases changes in representation:}
In the case of the cereal data set, students seemed to gain from re-using a well-known data set across exercises.
One could contrast this with Tufte's description of small multiples: 
`\textit{Constancy of design puts emphasis on changes in data, not changes in data frames}'~\cite{tufte1990envisioning}.
When learning about visualization techniques, perhaps constancy of data helpfully emphasizes changes in representation?
The specific value of the cereal data set seemed to be its large but manageable size, easy to understand semantics, and varied data types.

\textbf{Different data sets impose more time for students and the teaching team:}
The benefits of working with the cereal data set became crystal clear when students started working with the data sets they had collected on their own.
As teaching assistants, we started spending more time understanding their data structures and semantics, and how they had wrangled their data to suit their needs.
We think the data diversity also made it more difficult for students to engage in peer learning both within and between semester groups---suddenly, there was no one correct version of the data set.

On the other hand, allowing students to collect and work with their own data set seemed  motivating and empowering to students.
They grasped the opportunity to collect data about many different aspects of their life. Data sets included: exchanges of smiles with strangers, caffeine intake, sleeping patterns, use of emojis in text communication, photography habits, emotions and sleep, weather and mood, dietary habits and spending, screen time, listening behavior, clothing choices, types of (non-sexual) bodily touches, conversations and mood, meeting productivity, grocery shopping, alcohol consumption, student workload, and impact of medication on wellbeing.

To better manage the diversity of collected data, we consider providing a data template for future course iterations, perhaps one coupled with collection mechanisms such as provided by Google Forms and Sheets.
While such a template may not suit all possible student project needs, we expect this could drastically cut down on the difficulties students experienced in their projects.

We identify the following \textbf{course design goals for enabling diversity in what students \textit{do}:}
\textbf{Goal-4:} Critically assess when to provide concrete, specific direction to students and when and how to allow for flexibility.
\textbf{Goal-5:} Identify opportunities for scoping flexibility in ways that help students fulfill the intended learning outcomes, while getting smoothly through areas of project work that are less relevant for the course context.

\subsection{Communication between students, teaching assistants, and lecturers}
Communication is a cornerstone of teaching and learning---essentially, much of what we do \textit{is} communication. 
However, the communication that we want to discuss here is more specificly tied to visualization teaching, for example, as discussed above, through offering students to choose the data that they are working with. 


\textbf{The nearby meeting room provided a hub for communication and collaboration} 
We deferred questions that required consideration or dialog to office hours in the meeting room for both students and the teaching team (for example, questions that arose during or after lectures).
The easy access and the single place to defer to, made this a convinient option.
Thus, the room supported the teaching team in various communication needs---it became a place that teaching assistants would ask students to go with questions they were not able to answer on their own.

Many of the interactions with students in the meeting room related to their project work.
For example, we spent much time with individual student groups to align expectations.
Common questions were: `what is a sufficient level of interaction?', `how much design exploration is needed and how should we do it?', and `what level of design process articulation and reflection is expected?' 
While we had clarified many of these questions already during lectures and in the description of our expectations and grading approach for the project (see supplemental website), we saw these as expressions of anxious and stressed students that wanted to do well on the exam submission, although we frequently observed this at the expense of opportunities for learning.
We also spend much time with student groups to provide directions based on the concrete data that they were working with, potential visualization techniques to consider, and where to draw inspiration from.
Additionally, the meeting room served as a place for teaching assistants to check in with the course manager when in doubt about how to respond to a question, comment, or concern.
Further, it served as a place for commencing and summarizing the day within the teaching team.

\textbf{For electronic communication}, we shared materials via the course page, sent weekly summaries, and had the occasional e-mail exchange between students, groups of students, individual teaching assistants, groups of teaching assistants, the course manager, and the entire teaching team. 
This is all common practice.

The reflection hand-ins and weekly optional formative evaluation are less common practice, although the course manager routinely uses formative evaluation during courses.
However, in this course iteration, the reflection hand-ins (425 submissions) provided much stronger formative evaluation value than the Critical Incidents Questionnaire (31 submissions).
Moreover, the free-form nature of the 400-word reflections gave insights about what worked well and what needed improvement, for example, in the design exercises.
For example, it was through the reflection hand-ins, that we first became aware that the teaching team had different operationalisations of the 10 plus 10 approach.
The reflections also provided an overview of individual students ideas about their personal data visualization projects, which thus served as a vehicle to articulate and discuss some of these ideas in during lectures.

One might ask what these reflections are doing in a paper that is supposed to focus on visualization teaching and learning.
We find that our experiences in this course and reflections on them in relation to communication went beyond what we had seen before.
Further, we have not found similar discussions elsewhere.

Specifically, we want to discuss the visualization-specific aspects of our course communication.
Here, we do not mean specific to visualization in the topical sense, but specific in the sense of considering the state that visualization education is at---at least, the way that we see it.
As we noted previously, recent work has presented the current state of visualization education~\cite{bach_et_al:DagstuhlRep2022}.
In our context, we specifically considered the challenges brought about by growing student cohorts, which led us to reconsider the role of teaching assistants.

We started the course with the expectation that building a strong teaching team was important. 
This expectation came from experiences from previous iterations of the course, where recruiting teaching assistants was more difficult.
Following, this difficulty resulted in forming a teaching team that had only limited experience with the course and the topic.
For this iteration however, we were able to compose a strong teaching team that already knew each other and the lecturer well (a great precondition for good communication in the teaching team).
As the course progressed, we became increasingly aware of our use of a range of communication channels and modalities and of the importance of good and stable communication.

We
%
%
identify the following \textbf{course design goals for communication:} 
\textbf{Goal-6:} Allow for experimentation in teaching and learning and to include teaching assistants in doing so.
\textbf{Goal-7:} Disseminate teaching and facilitation skills, knowledge, and practices from people trained in this area (such as lecturers) to teaching assistants and other members of visualization teaching teams.
\textbf{Goal-8:} Build strong communities of practice~\cite{wenger1998communities} framed specifically around teaching assistants and forming their role in large visualization courses.

\section{Future work}
Here, we briefly discuss future work in terms of our local context and practice, as well as future research opportunities.

Writing this, we are about six months out from starting the next course iteration.
Thus, we have begun to discuss \textbf{concrete changes to course structure and content}.
While we expect to keep the design process structure in the semester, we plan to experiment with a weekly flip-flop between design and programming exercises.
This is motivated mainly by two considerations. 
First, we observed that the 2023 iteration introduced technical tools necessary for the students' project quite late in relation to their project deadlines.
Second, we hope that students with stronger design or programming skills will be better able to use these as scaffolds to learn the other course elements.
We also think an earlier introduction to technical tools will help students better frame their project work in data, provide concrete tools useful for data processing, and enable a more designerly approach to visualization programming.
We also expect to move \textbf{from visualization specification towards visualization programming}.
Students in our program are already introduced to JavaScript in a course running parallel to the visualization course.
Additionally, there are plans to replace Python with JavaScript in the introductory programming course.
Thus, we plan to introduce D3.js and a more exploratory programming approach in the next course iteration.
This change will also facilitate visualization products that are embedded in a clearer contexts of use and affording increased room for design variations, compared to Google Colab or similar notebook formats.
Moreover, we consider a \textbf{stronger emphasis on data wrangling aspects}, based on feedback in response to design briefs and in-person supervision in the meeting room.
Based on this, we plan to include literature and spend time during lectures to cover topics in relation to this (such as data wrangling). 
To help students more easily get on track in their data collection, we also consider providing them with a data collection template.
However, this decision is pending a proper teaching team discussion with focus on intended learning outcomes.

The present paper provided limited accounts of student materials, results, examples, and anecdotes.
Since we did not request ethics consent from students to include their work in this paper, we were unable to do so.
We plan to obtain informed consent to share materials from future iterations of the course, as well as plans for inviting students from the 2023 course iteration to provide post-hoc consent to the use of student materials, to respond to a questionnaire, and to participate in a focus group. 
Thus, while including student materials, results, and anecdotes is beyond the scope of this paper, we hope to be able to follow up on these aspects in the future. 

Finally, the course design goals that we identified in Section \ref{sec:themes} can be readily considered as research challenges and map well to other visualization education challenges~\cite{bach2023challenges}. For example, in relation to \textbf{Goal-7}, future research might seek identify good ways to enable knowledge dissemination with visualization teaching teams. 

\section{Conclusion}
We contributed a rich account of six design activities in terms of purpose, structure, and outcome in addition to their relation to subsequent programming activities and an overall course context.
From this perspective, we presented three themes that emerged from reflections between members of the teaching team.
First, we reflected on scaffolding in our course design and the relevance of continuous planning.
Through this, we uncovered difficult moments in terms of scaffolding and shared our reflections on possible ways to address these challenges.
Second, we reflected on diversity in student projects and activities, with a strong focus on data.
Third, based on the first two themes, we reflected on the communication between students, teaching assistants, and lecturers, and how our choices of communication related to the other themes. 
Doing so, our reflections form a basis for fellow visualization educators for designing visualization teaching and learning.


\acknowledgments{%
We wish to thank Nynne Dyre, who worked with us as teaching assistant as part of our teaching team as well as the students who attended the 2023 data visualization course. 
We also thank the many people who have influenced our teaching approaches in this course.
Most importantly, Professor Sheelagh Carpendale shaped our approach to teaching data visualization through hands-on approaches.
}

\bibliographystyle{abbrv-doi-hyperref}

\bibliography{template-manual,template-zotero}


\appendix
\onecolumn

\end{document}